\documentclass{iau} 
\usepackage{graphicx}

\title[Merging massive BH] 
{Merging massive black holes \\ the right place and the right time}

\author[Astrid Lamberts]   
{Astrid Lamberts$^1$}

\affiliation{$^1$ Theoretical Astrophysics, California Institute of
  Technology,  \\ Pasadena, CA, 91125, USA
\\ email: {\tt lamberts@caltech.edu}} 

\pubyear{2018}
\volume{338}  
\jname{Gravitational Waves Astrophysics}
\editors{A.C. Editor, B.D. Editor \& C.E. Editor, eds.}
\begin{document}

\maketitle

\begin{abstract}
The LIGO/Virgo detections of gravitational waves from merging black holes of  $\simeq$ 30 solar mass
suggest progenitor stars of low metallicity
($Z/Z_{\odot} \lesssim 0.3$). In this talk I will provide
constrains on where the progenitors of GW150914 and GW170104 may have
formed, based on advanced models of galaxy formation and evolution
combined with binary population synthesis models. First I will combine
estimates of galaxy properties (star-forming gas metallicity, star
formation rate and merger rate) across cosmic time to predict the low
redshift BBH merger rate as a function of present day host galaxy
mass, formation redshift of the progenitor system and different
progenitor metallicities. I will show that the signal is dominated by
binaries formed at the peak of star formation in massive galaxies with
and binaries formed recently in dwarf galaxies. Then, I will present
what very high resolution hydrodynamic simulations of different galaxy
types can learn us about their black hole populations.

\end{abstract}

\firstsection 
\section{Introduction}
The detection of gravitational waves (GW) from merging black holes (BH) and
neutron stars has opened a new window on our Universe (\cite{GW150914,BNS}). More specifically, it has revived the study of massive stars and their binary interactions (\cite{Belczynski16,Eldridge16,Spera}). While there is a global understanding of binary evolution for field binaries, the exact properties of their mass loss, mass transfer and supernova mechanisms are still unclear (\cite{Belczynski10,Dominik13}). The high masses of the first GW detection, (30 $M_{\odot}$ BHs for GW150914) were unexpected (but see \cite{Belczynski10B}). These high masses suggest low-metallicity progenitors, where mass loss due to winds is reduced. We have limited information on low-metallicity massive stars as massive stars are short-lived and stars recently formed in the Milky Way have a metallicity at least as high as the Sun. In this talk, we determine the conditions where the progenitors of GW150914 may have formed, based on a semi-analytic formalism of galaxy evolution and a binary population synthesis model. We determine the formation rate of GW150914-like progenitors based on the mass of the host galaxy and the formation time of the progenitor. We refer the reader to \cite{Lamberts16} for a more detailed description of the method.  We then present a preliminary analysis of the merger rates predicted by cosmological simulations.

\section{Forming low-metallicity stars}
\begin{figure}[h]
\begin{center}
 \includegraphics[width=2.2in]{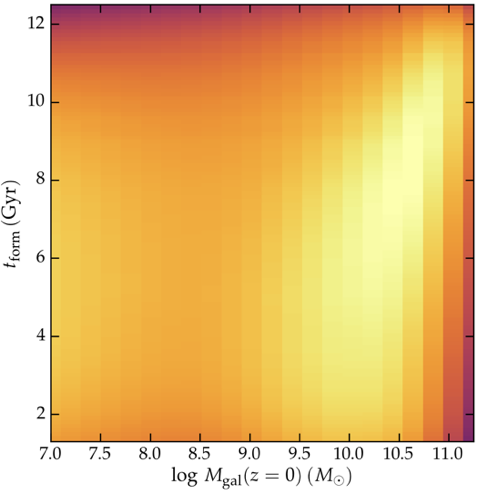} 
 \includegraphics[width=2.3in]{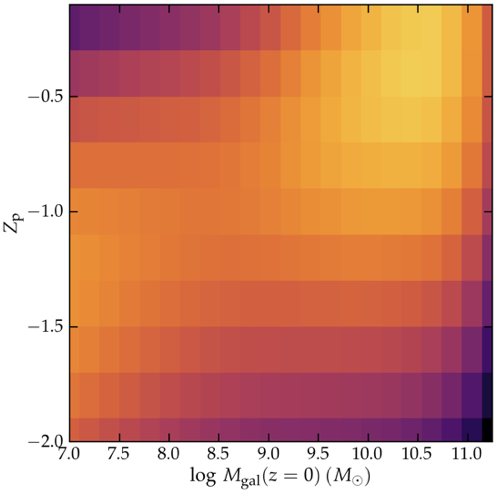}
 \caption{Low-metallicity star formation rate (arbitrary units) as a function of galaxy mass and lookback time to  formation (left) and metallicity (right).  }
   \label{SFR}
\end{center}
\end{figure}

The first step is to determine the conditions for low-metallicity star formation in the universe. Globally, the metallicity in the Universe increases as successive generations of stars progressively enrich the star-forming gas. At a given epoch galaxies follow the mass-metallicity relation (\cite{Kewley08}), with more massive galaxies being more metal rich.  For each present-day galaxy mass, we determine the mass of the corresponding dark matter halo using abundance matching by  \cite{Behroozi13}. Based on the model by \cite{Behroozi13} we then determine the amount of star formation in each halo in 10 Myr time bins.  Based on the redshift-dependent mass-metallicity relation computed by \cite{Ma16}, we derive the stellar mass formed  in 11 metallicity bins between $Z=0.01Z_{\odot}$ and  $Z=1Z_{\odot}$, where $Z_{\odot}\equiv 0.02$.  We include scatter between different galaxies of $\sigma=0.1$ dex, based on observations by \cite{Tremonti04}. We also include $\sigma=0.2$ scatter of the metallicity within galaxies, according to \cite{Berg13}. Fig.~\ref{SFR} shows the normalized star formation rate as a function of present-day galaxy mass and lookback time to formation (left) and metallicity (right). The figure only shows subsolar metallicity star-formation, which is relevant to binary BH formation. However, particularly in massive galaxies, a large fraction of the stars form at higher metallicities. We find a slightly bimodal distribution, with most of the low-metallicity stars forming in massive (Milky-Way like) galaxies around the peak of cosmic star-formation (between 6 and 10 Gyrs ago). More recent low-metallicity star formation also occurs in dwarf galaxies, typically at metallicity below 10 per cent of Solar. In the next section, we will highlight how metallicity affects binary BH formation and merger rates and show which low-metallicity stars shown in Fig.~\ref{SFR} mostly contribute to GW150914-like events observed now.

\section{Forming binary black hole merger progenitors}

\begin{figure}[h]
\begin{center}
 \includegraphics[width=3.4in]{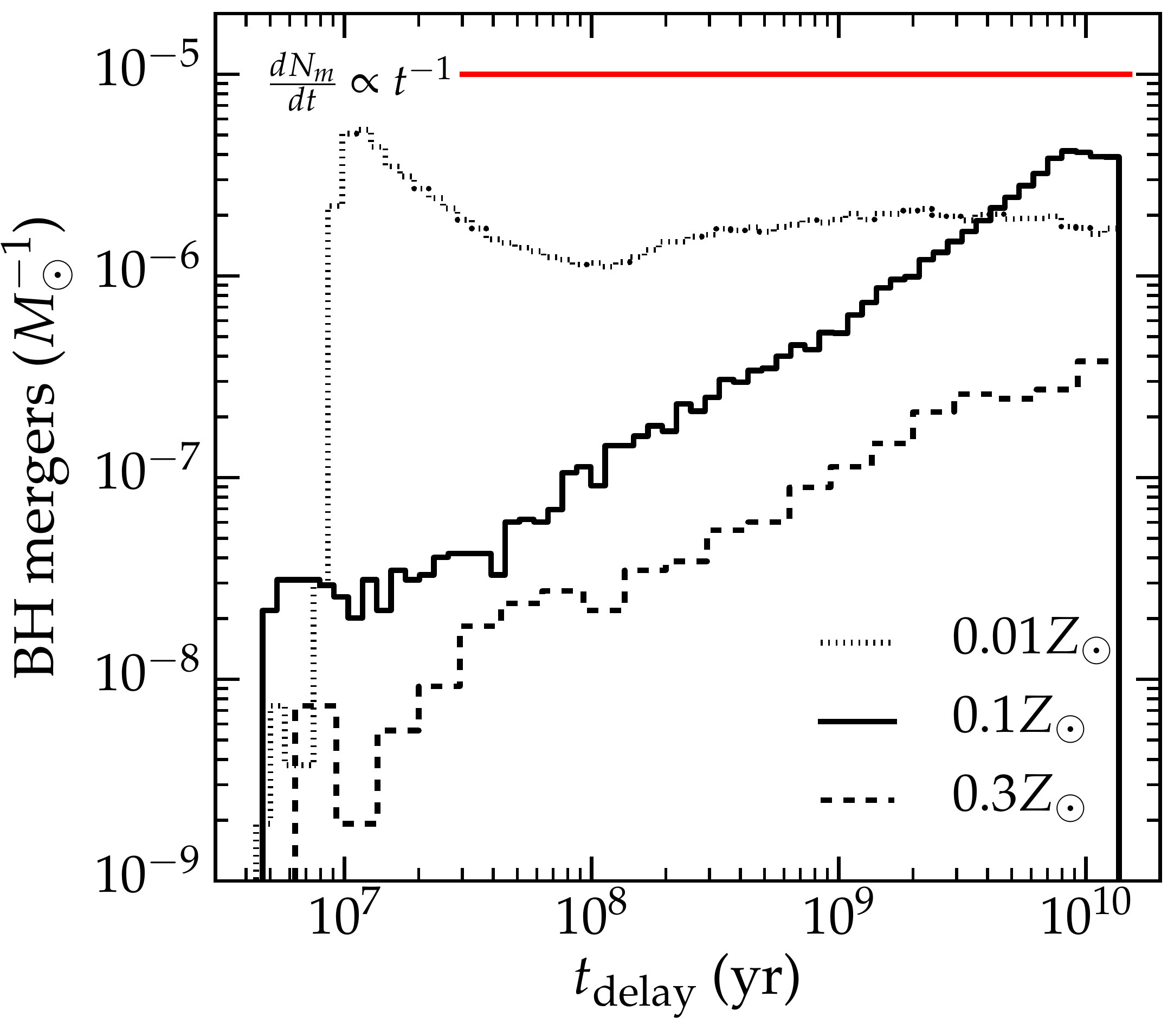} 
 \caption{Delay time distribution for massive binary black hole mergers with progenitors of different metallicities. Taken from \cite{Lamberts16}. }
   \label{delay}
\end{center}
\end{figure}

We use the BSE code (\cite{Hurley02}) to compute the delay time distribution of binary BH mergers  with respect to the formation of the stellar progenitor. We neglect BH mergers from globular clusters (\cite{Rodriguez15}) and consider a single set of standard assumptions on binary evolution. We have updated BSE to account for improvement models of stellar winds (\cite{Belczynski10}), remnant masses (\cite{Belczynski08}) and BH kicks due to supernova explosions (\cite{Dominik13}). Binaries undergoing common-enveloppe mass-transfer during the Hertzprung gap merge as stars. At later stages, we set the common enveloppe efficiency to unity and assume that during Roche lobe overflow, only half of the mass is accreted by the companion. For each metallicity bin, we model $2.5\times 10^6$ binaries with primary masses between 25 and 150 $M_{\odot}$ following the \cite{Kroupa01} initial mass-function. The initial period distributions and mass ratios are derived from observations by \cite{Sana12} and we use  thermal distribution for the initial eccentricity. Fig.~\ref{delay} shows the BH mergers with total mass above 40 $M_{\odot}$ (per unit solar mass) for $Z=0.01, 0.1,0.3 Z_{\odot}$ according to our models. The lowest metallicity stars produce the most massive BH, as their mass-loss is limited and early mergers. At later times, more metal-rich stars contribute more. Stars with metallicity above 0.3 Z$_{\odot}$ produce 10 times less mergers of these high masses, owing to the strong mass loss of the progenitor stars.

\begin{figure}[h]
\begin{center}
 \includegraphics[width=2.2in]{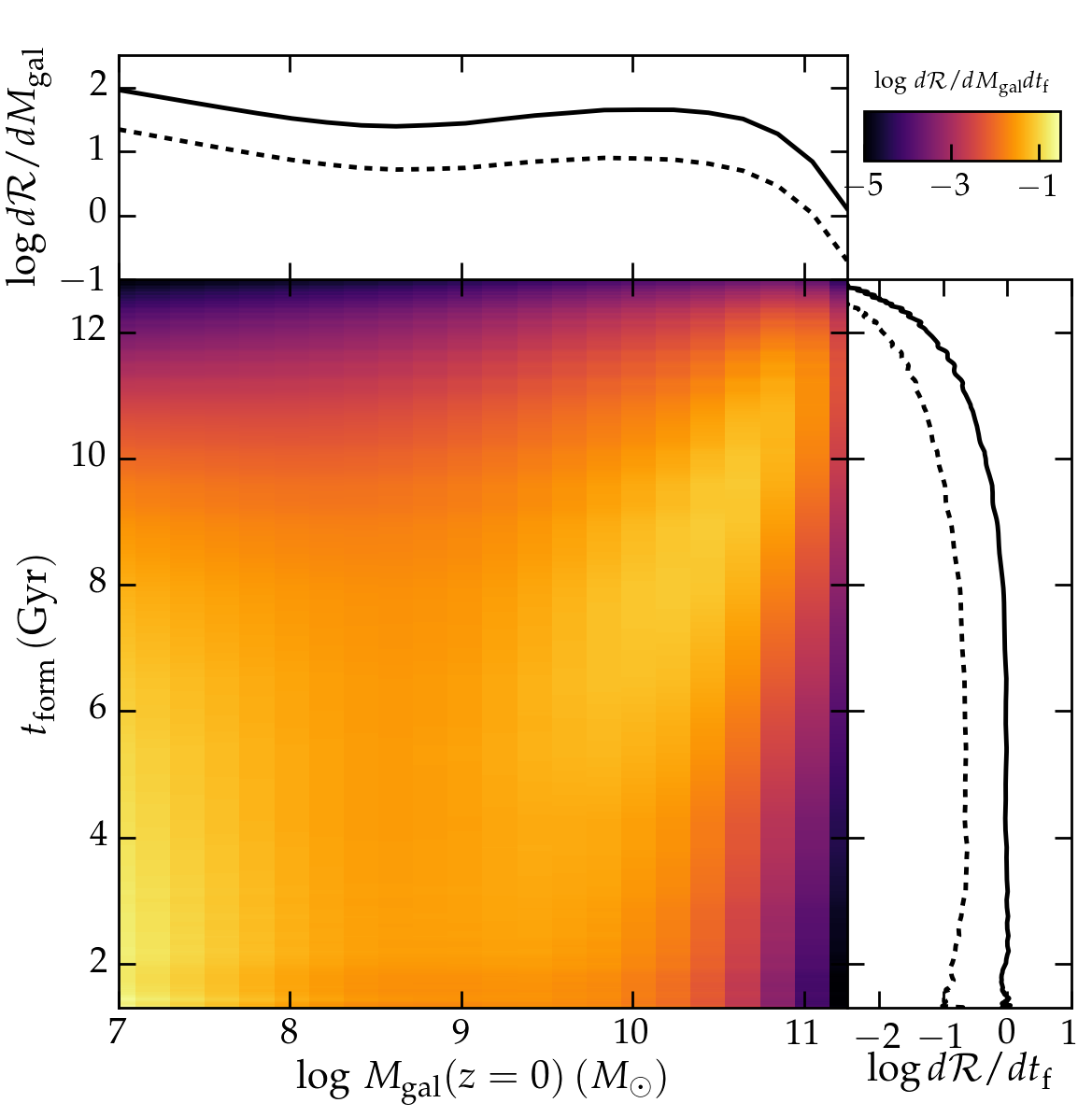} 
 \includegraphics[width=2.3in]{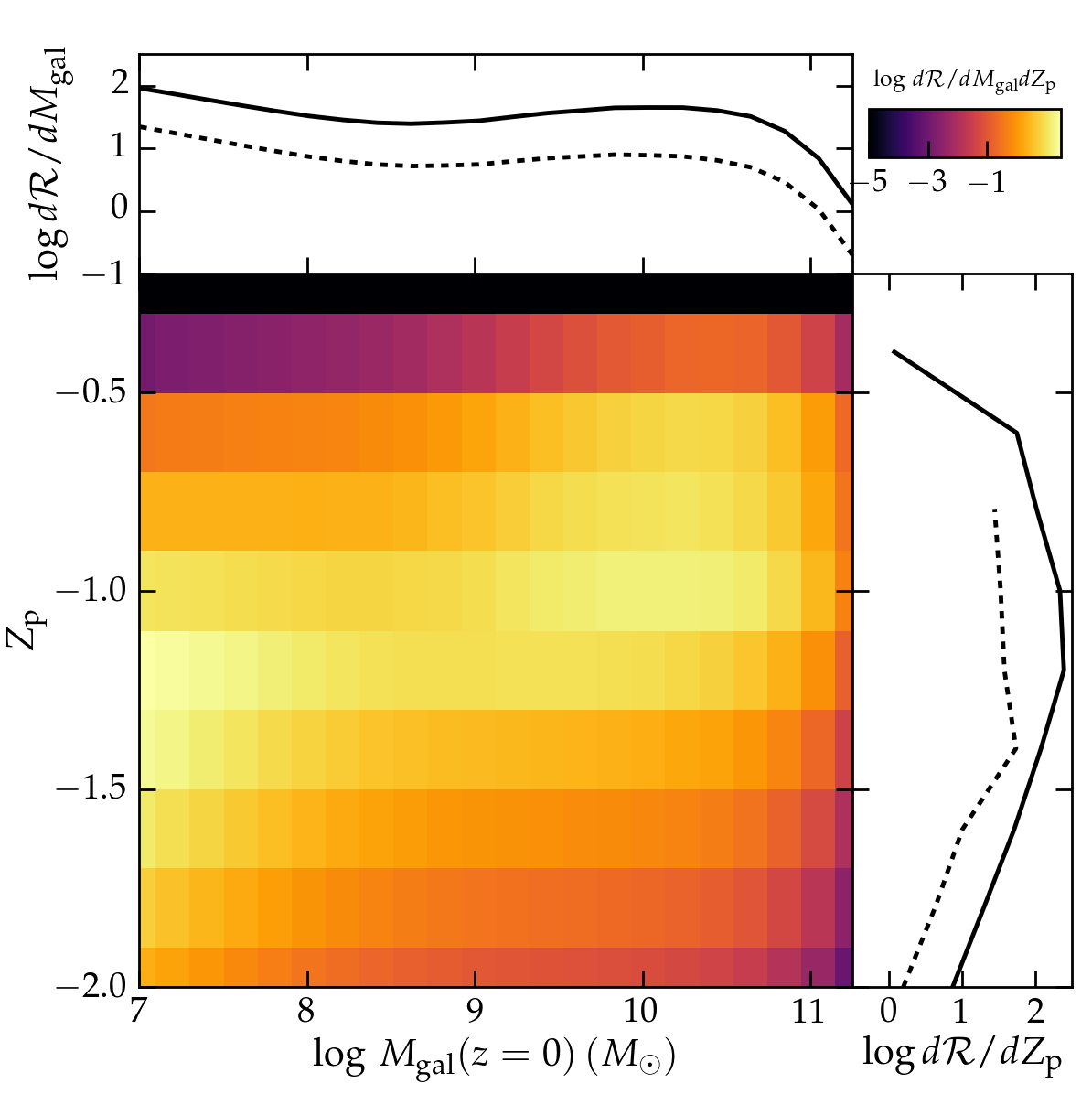}
 \caption{ Formation rate of progenitors of massive black hole mergers as a function of galaxy mass and lookback time to formation (left) and metallicity (right). Taken from \cite{Lamberts16}. }
   \label{merger_formation}
\end{center}
\end{figure}

We combine our semi-analytic model for low-metallicity star-formation (Fig.~\ref{SFR}) and the metallicity-dependent delay-time distribution (Fig.~\ref{delay}) to determine where and when the progenitor of GW150914 most likely formed. For all stars formed, at all timesteps, we count the resulting BH mergers occurring between $z=0.1$ and the present day.  The conditions of formation of the progenitors are shown in Fig.~\ref{merger_formation}. The figure has the same axes as Fig.~\ref{SFR} and the colors can be directly compared. It shows that the bimodality in low-metallicity star formation is enhanced. Roughly half of the progenitors of GW150914-like systems stem from Milky-Way like galaxies ($M\simeq 10^{11}M_{\odot}$) and were formed 6 to 10 Gyrs ago, at roughly 10 per cent of Solar metallicity.  Later star formation in massive galaxies has too high metallicity to significantly contribute to BH merger progenitors. Recent star formation in dwarf galaxies also significantly contributes to BH mergers, stemming from progenitors with metallicities below 10 per cent of Solar.  Because of their low-metallicity, galaxies below $10^8 M_{\odot}$ overproduce BH mergers in comparison with their contribution to the global star formation.  These galaxies are faint, and unobservable at high redshift and BH mergers may be the only way to infer some of their properties. The side panel of the left figure shows the formation time of the progenitors of massive BH mergers, which is mostly uniform over the past 7 Gyrs. This differs from the bimodal distribution from \cite{Belczynski16}, because of our improved model of low-metallicity star-formation. In the next section we present preliminary results from cosmological simulations.

\section{Preliminary results from high-resolution simulations}

\begin{figure}[h]
\begin{center}
 \includegraphics[width=2.3in]{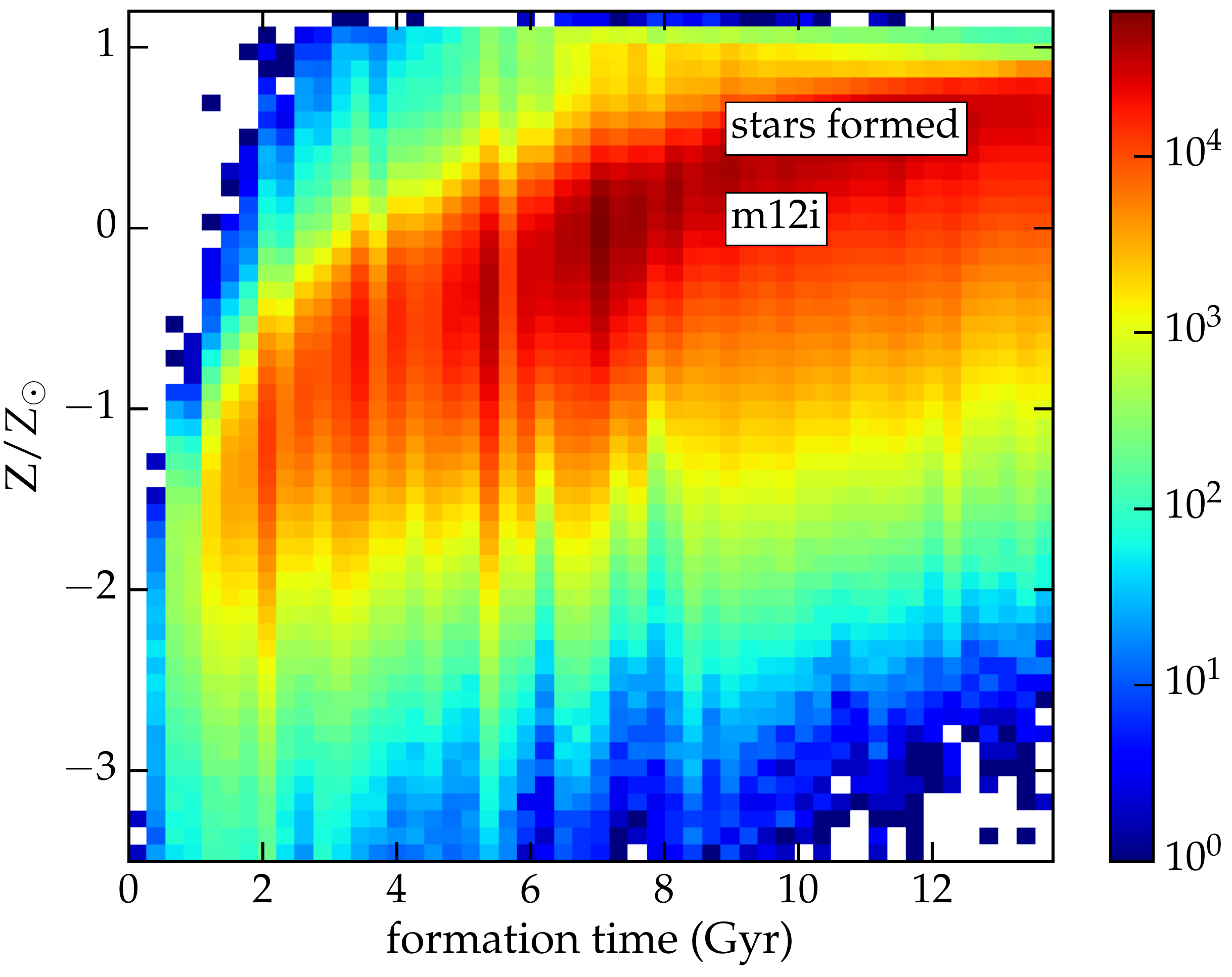} 
 \includegraphics[width=2.3in]{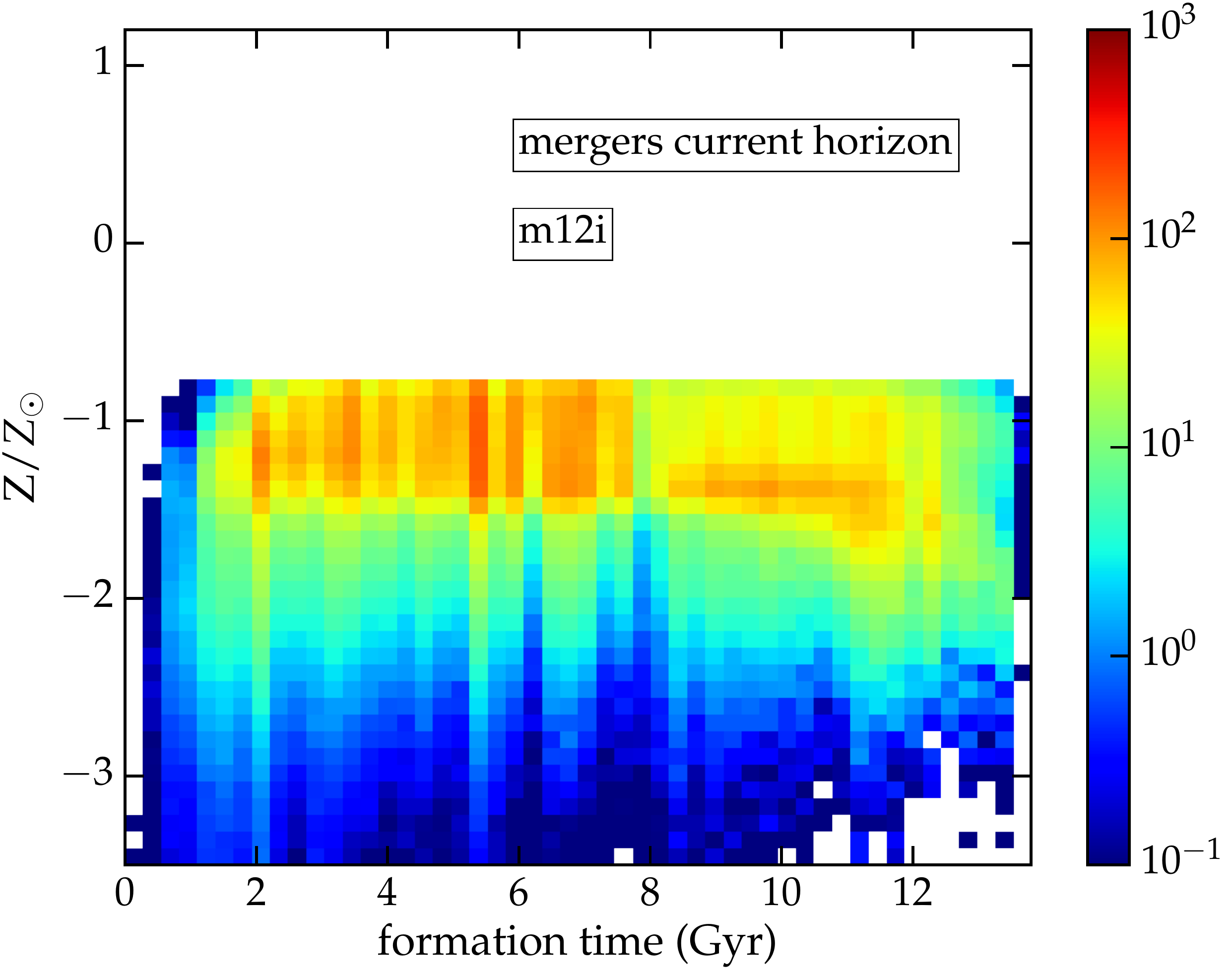}
 \caption{Normalized star formation rate (left) and formation rate of BBH merger progenitors (right) from the Latte simulation. }
   \label{Latte}
\end{center}
\end{figure}

\begin{figure}[h]
\begin{center}
 \includegraphics[width=2.2in]{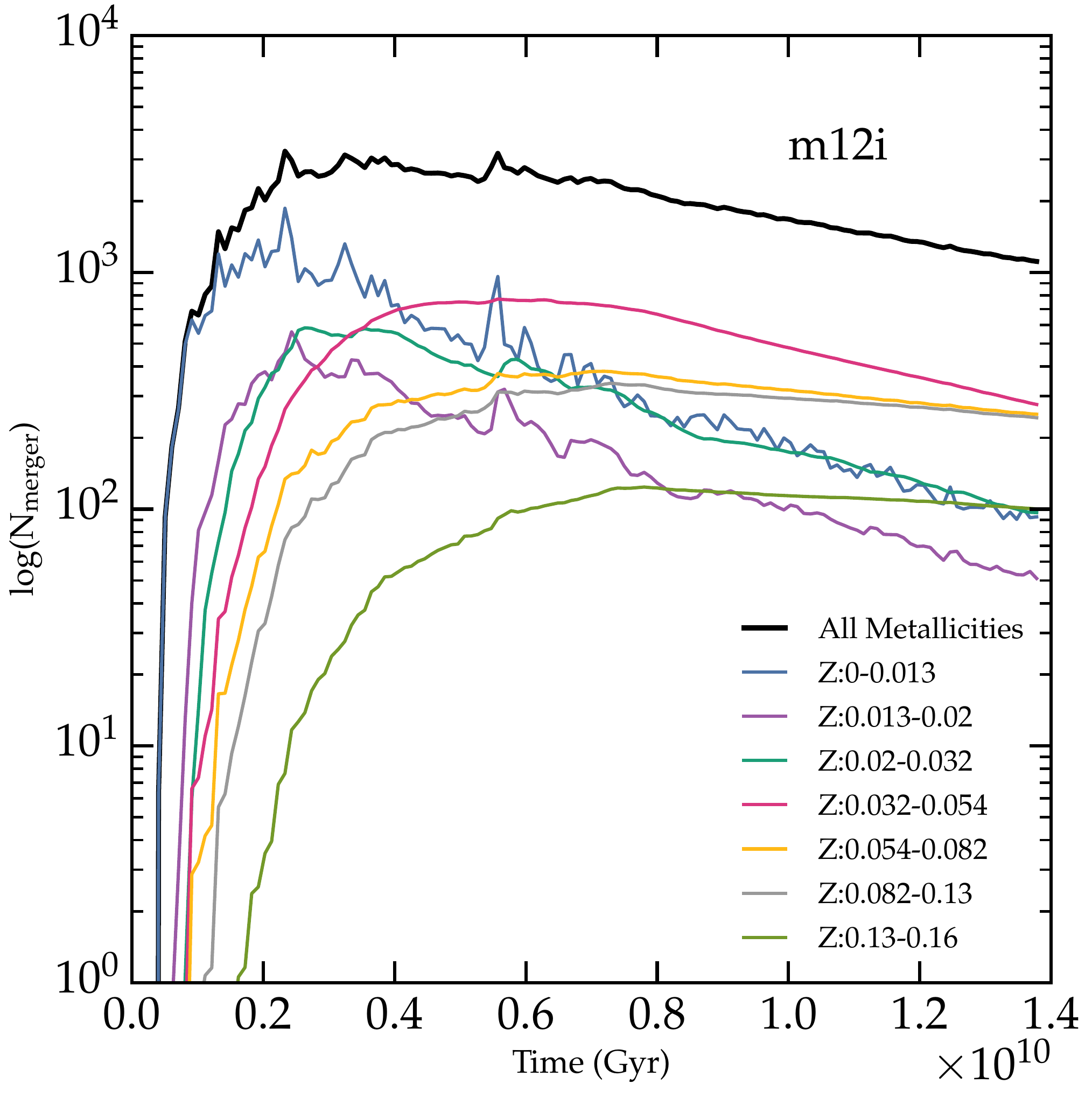} 
 \caption{Number of mergers in a Milky-Way like galaxy as a function of time for different metallicities of the progenitor stars. The total number of mergers is shown in the thick black line.}
   \label{all_mergers}
\end{center}
\end{figure}

We have performed  a comparable analysis based on hydrodynamical simulations of different galaxies.  We use simulations from the Feedback In Realistic Environments project (FIRE ,\cite{HopkinsFIRE}). The cosmological simulations are ran with the GIZMO code (\cite{HopkinsGIZMO}) and include  a multiphase model of the interstellar medium and stellar feedback based on a complete stellar evolution model. We specifically use the ``Latte'' simulation (\cite{Wetzel16}), a cosmological simulation of a Milky-Way like galaxy with a mass resolution of 7070 $M_{\odot}$. The left panel of fig.~\ref{Latte} shows the normalized star-formation rate in a Milky-Way like galaxy. The right panel only shows the stars contributing to BH mergers within the detector horizon during the second observing run. We have directly associated the delay time distributions (Fig.~\ref{delay}) with the star formation in the simulation. We find that mergers come from low-metallicity star-formation and that effectively most of the stars formed in Milky-Way-like galaxies are unable to form merging black holes. As in \cite{Lamberts16}, we find that the mergers come from star formation between 6 and 10 Gyrs ago.  Fig.~\ref{all_mergers} shows the total number of mergers as a function of time in different metallicity bins. Again, we confirm the findings from \cite{Lamberts16}, with the largest contribution coming from $Z=0.03-0.13 Z_{\odot}$. We find that the number of mergers increases towards redshift $z=2$ (10Gyrs ago), somewhat following the global star formation history of the galaxy. As the horizon of the detectors increases, we thus expected an increase of the number of detections per unit volume for contributions from massive galaxies.

\section{Implications}
In this talk, we presented a unique combination between a binary population synthesis model and a complete semi-analytic cosmological model for low-metallicity star formation. We derive the contribution of different galaxies to the present-day merger rate of massive ($>30 M_{\odot}$) binary black holes. We find that the progenitors of such mergers could have formed during the peak of star formation (between 6 and 10 Gys argo) in a galaxy that would now resemble the Milky Way or more recently ($<6$ Gyrs ago) in a dwarf galaxy. Our method solely relies on the strong metallicity dependence of binary black hole formation (and mergers) to determine the \textit{relative} contributions of different galaxies to present-day black hole mergers.  Although \textit{absolute} merger rate we predict is in agreement with the LIGO/Virgo detection rate (\cite{LIGOrate}), we emphasize that it may be revised as our understanding of massive binary evolution improves.  We confirm our findings using the same binary evolution model combined with a high-resolution simulation of a Milky-way like galaxy. We find that the merger rate in massive galaxies increases towards redshift $z=2$ and hundreds of detections are expected when LIGO and Virgo reach their design sensitivity (about $z\simeq1$). As such, these detections will provide strong constrains on binary evolution models and star formation in high-redshift and/or faint galaxies.

\end{document}